\documentclass[sigconf, nonacm]{acmart}

\newcommand\vldbdoi{XX.XX/XXX.XX}

\newcommand\vldbavailabilityurl{https://github.com/AlgazinovAleksandr/Multi-Agent-MATE}

\usepackage{amsmath}
\usepackage{amsfonts}
\usepackage{algorithmic}
\usepackage{graphicx}
\usepackage{textcomp}
\usepackage{xcolor}
\usepackage{multirow}
\usepackage{listings}
\usepackage{caption}
\usepackage{hyperref}
\usepackage{subcaption}

\begin{document}
\title{MATE: LLM-Powered Multi-Agent Translation Environment for Accessibility Applications}

\author{Aleksandr Algazinov}
\affiliation{%
  \institution{Tsinghua University}
  \city{Beijing}
  \country{China}
}
\email{algazinovalexandr@gmail.com}

\author{Matt Laing$^*$}
\affiliation{%
  \institution{Tsinghua University}
  \city{Beijing}
  \country{China}
}
\email{matthieu.laing@gmail.com}

\author{Paul Laban$^*$}
\affiliation{%
  \institution{Tsinghua University}
  \city{Beijing}
  \country{China}
}
\email{plaban.pro@gmail.com}

\thanks{$^*$ - Equal contribution}

\begin{abstract}
Accessibility remains a critical concern in today's society, as many technologies are not developed to support the full range of user needs. Existing multi-agent systems (MAS) often cannot provide comprehensive assistance for users in need due to the lack of customization stemming from closed-source designs. Consequently, individuals with disabilities frequently encounter significant barriers when attempting to interact with digital environments. We introduce MATE, a multimodal accessibility MAS, which performs the modality conversions based on the user's needs. The system is useful for assisting people with disabilities by ensuring that data will be converted to an understandable format. For instance, if the user cannot see well and receives an image, the system converts this image to its audio description. MATE can be applied to a wide range of domains, industries, and areas, such as healthcare, and can become a useful assistant for various groups of users. The system supports multiple types of models, ranging from LLM API calling to using custom machine learning (ML) classifiers. This flexibility ensures that the system can be adapted to various needs and is compatible with a wide variety of hardware. Since the system is expected to run locally, it ensures the privacy and security of sensitive information. In addition, the framework can be effectively integrated with institutional technologies (e.g., digital healthcare service) for real-time user assistance. Furthermore, we introduce ModCon-Task-Identifier, a model that is capable of extracting the precise modality conversion task from the user input. Numerous experiments show that ModCon-Task-Identifier consistently outperforms other LLMs and statistical models on our custom data. Our code and data are publicly available at \href{https://github.com/AlgazinovAleksandr/Multi-Agent-MATE}{https://github.com/AlgazinovAleksandr/Multi-Agent-MATE}.
\end{abstract}

\maketitle



\section{Introduction}

Multi-Agent Systems (MAS) are increasingly applied across diverse domains such as autonomous driving, healthcare, and finance \cite{sun2025multiagentcoordinationdiverseapplications}. For instance, in autonomous vehicles \cite{autonomous_driving}, MAS facilitate decentralized traffic coordination, improving the safety and efficiency of urban traffic light control and vehicular flow in real-time \cite{9103316}. In healthcare, agents collaborate on patient monitoring, diagnostics, and resource management. Within smart grids \cite{mas_smart_grid}, MAS manages demand-response operations and optimizes energy distribution. In logistics and search and rescue \cite{mas_sar}, agents divide search, rescue, and coordination tasks, make quick decisions about route planning, resource allocation, or mission priorities, and react to new problems like blocked paths or accidents, enhancing operational resilience. \cite{tran2025multiagentcollaborationmechanismssurvey}. The advantages of MAS include having no single point of control, being able to keep working even if one part fails, and easily adding more or adapting existing agents if and when needed, making them suitable for dynamic, real-world environments \cite{tran2025multiagentcollaborationmechanismssurvey}. Each agent is capable of both autonomous decision-making and collaboration, resulting in more effective problem-solving. These features are important in systems that are required to adapt to uncertainty, like noisy data, changing surroundings, or missing information. Hence, MAS are crucial for enhancing complex systems such as autonomous transportation networks, smart grids, and financial trading systems \cite{smartcities5010019}. \\

People with disabilities often face challenges when using technology, especially when their specific needs are not being met. Disabilities can affect how people move, see, hear, or think, so they may require different ways to interact with digital tools. However, many everyday technologies, especially MAS, are not flexible enough to satisfy their needs, making basic tasks, communication, or daily life harder and more uncomfortable for them. Although MAS are effective in numerous applications, their use in accessibility is still limited. Most MAS implementations for accessibility are typically implemented for specific tasks, such as voice control tools or screen readers for blind users, and cannot be generalized over various accessibility use cases \cite{tran2025multiagentcollaborationmechanismssurvey}. Moreover, most of the existing solutions are not open source and, therefore, not customizable and widely accessible \cite{smartcities5010019}. These limitations highlight the necessity for a general-purpose, open-source, and lightweight MAS architecture that allows for real-time, multi-modal adaptation to different people's accessibility needs and requirements. \\

To address these challenges, we introduce MATE, a comprehensive multi-agent AI accessibility system that uses advanced AI tools and a complex code structure to assist users in need. \textbf{Our contributions are summarized as follows:}

\begin{itemize}

    \item We introduce MATE, a comprehensive, flexible, and open-source MAS that understands and solves multiple modality conversion problems based on the user's needs. To the best of our knowledge, this is the first open-source and lightweight MAS built specifically for modality adaptation tasks. The most closely related open-source solutions include general-purpose MAS \cite{openmanus2025} \cite{owl}, and Multimodal Large Language Models (MLLMs) designed to effectively handle different modalities \cite{Ming-Omni} \cite{Next-GPT} \cite{LLMBind}. \\

    \item We created the ModConTT (Modality Conversion Task Type) dataset. The dataset was AI-generated and human-verified. Verifications confirmed the correctness, completeness, and diversity of the dataset, making the dataset a suitable benchmark for further research within the AI for accessibility domain. \\

    \item We introduce ModCon-Task-Identifier, a model designed for recognizing the modality conversion task type based on the user prompt. Furthermore, our experimental findings show that the model significantly outperforms other existing LLMs (e.g., Llama-3.1-70B-Instruct), as well as machine learning classifiers (e.g., CatBoost \cite{CatBoost}) trained on the ModConTT dataset. The models were evaluated using standard classification metrics, such as accuracy and F1-score.
    
\end{itemize}

\section{Related Work}

\begin{figure*}[htbp]
    \centering
    \includegraphics[width=1\linewidth]{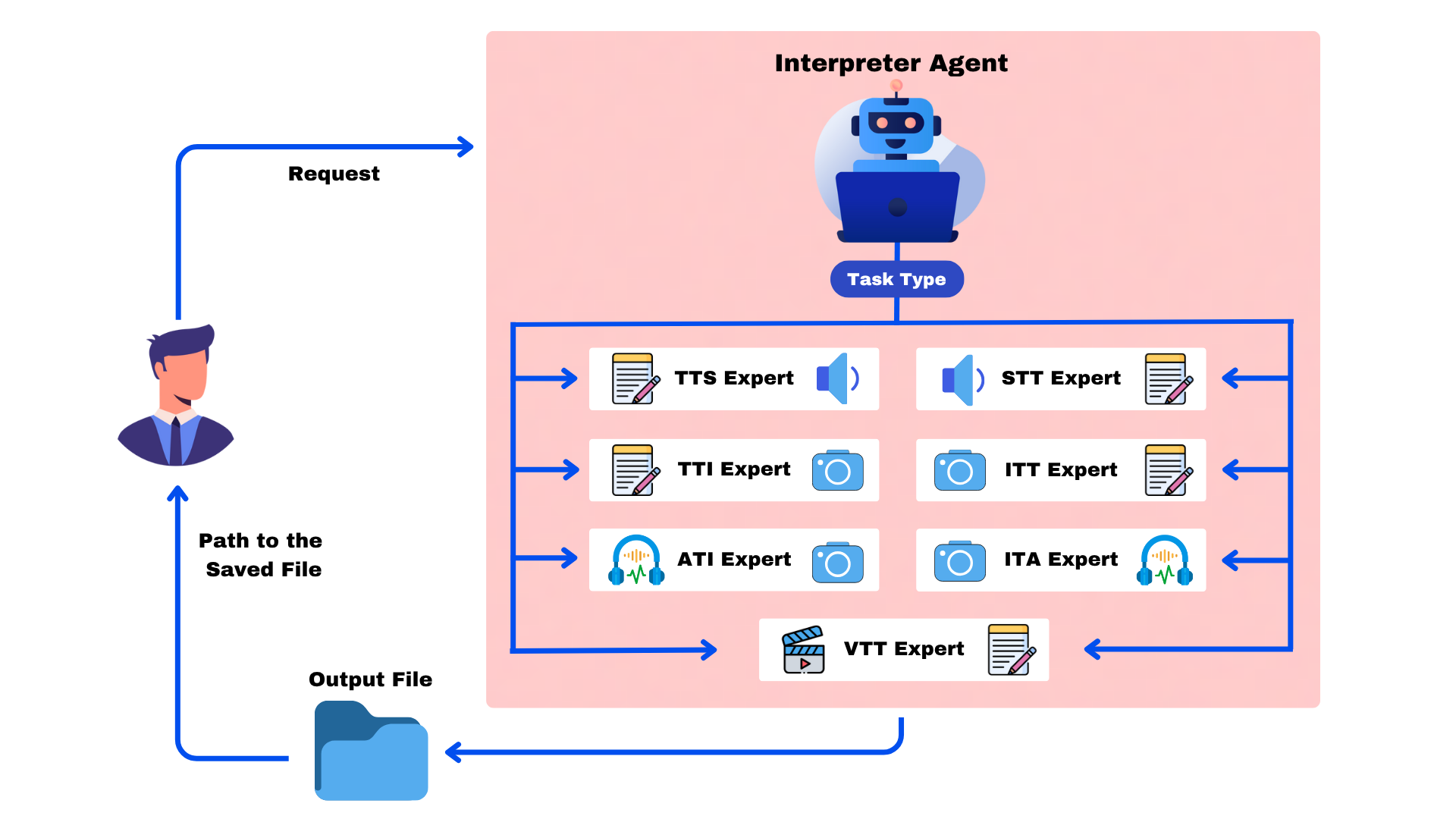}
    \caption{MATE Workflow}
    \label{fig:architecture}
\end{figure*}

\subsection{Multi-Agent Systems}

In recent years, MAS has proven to be a powerful tool for solving problems for which single-agent AI solutions are inefficient \cite{LLM-vs-MAS}. Because each agent in the system can be specialized for a single task, those systems are more relevant and have more adaptability when in complex environments. Numerous studies showed that the use of MAS was effective on a wide variety of tasks \cite{LLM_Agents} \cite{AutoGen}, including decision-making, reasoning, etc. Furthermore, the efficiency and saved time of dividing work across multiple agents are essential in applications such as numerical accessibility. Recent advances in multimodal learning \cite{CAML} have made MAS more effective than single-agent systems in dynamic environments. However, agents require efficient communication as well as precise prompts in the context of LLMs. Studies \cite{Mutli_Agents_Fail} have shown that MAS lacking those elements are inferior to single-agent models in problem-solving skills. The same studies proposed multiple strategies to improve MAS, such as establishing a standardized communication protocol and reinforcing clear role specifications by defining conversation patterns and setting termination conditions. Another drawback of a multi-agent system context is the scale needed for the whole solution, particularly when using LLMs. This issue is addressed by new frameworks aiming to improve small language models and their reasoning skills \cite{Improving_Function_Calling} \cite{ToolLLM} \cite{ReAct}, thus reducing the necessary size needed in MAS. Recent studies integrate MAS with Large Language Models (LLMs), boosting their cognitive flexibility, reasoning, planning, and adaptability. Frameworks like AutoGen \cite{AutoGen} and LangChain \cite{langchain} demonstrate how LLM-powered agents can engage in collaborative multi-agent dialogues and tool use \cite{ToolLLM}, simulating expert interaction. These integrations mark a shift from rule-based coordination to adaptive, knowledge-driven agent collaboration \cite{tran2025multiagentcollaborationmechanismssurvey}. These developments underscore MAS as fundamental architectures for scalable, intelligent, and multimodal AI ecosystems that support emergent intelligence in human-AI collaboration \cite{sun2025multiagentcoordinationdiverseapplications}.

\subsection{AI for Accessibility}

Numerous AI tools were built to improve accessibility for users with disabilities  (e.g., real-time captioning, content description, etc.). The primary objective is to ensure that people with disabilities can exploit AI capabilities to enhance their learning or social interactions. AI is becoming more and more important in enhancing cognitive support, mobility, and medical diagnostics, allowing for user-specific and flexible solutions \cite{Multimodal_Cognitive_Support} \cite{Foundation_to_Assist}. Thus, many tools are developed to fit different types of conditions, such as natural language processing (NLP) for speech recognition and translation, open models for text-to-speech, or even sign language transcription to support interaction innovation \cite{AI_Accessibility} \cite{Evaluating_Multimodal_Visual_Assistants} \cite{AI_Healthcare} \cite{Digital_Accessibility}. In addition, smart homes and tools are produced to help users (mostly elderly people), assisting them daily. In this paper, we will compare our work to research based on multi-agent interaction \cite{Multi-Agent_Assist} that reaches multiple limitations since it is only focused on visually impaired disabilities with a single camera that captures information to be treated by the multi-agent. Thus, our research targets a larger audience with diversified disabilities that can be treated and handled by MAS.

\subsection{Modality Conversion}

Modality conversion is defined as mapping data from one form to another. This is an important problem in the AI for accessibility domain, since the goal is to represent the particular data in a way that is convenient for a user. Some of the modality conversion tasks can be summarized as follows:

\begin{itemize}

    \item \textbf{Text-to-Sign Language (TTSL)}. As of 2021, sign language (SL) was used by 466 million people as the main source of communication \cite{How2Sign}. Techniques for solving the problem include Text-to-Gloss-to-Pose-to-Video \cite{Sign-Language-Production} (a combination of Text-to-Gloss, Gloss-to-Pose, and Pose-to-Video models) and gloss-free approaches \cite{SignGen}. While numerous models for different sign languages exist \cite{Sign-IDD} \cite{Gloss-Based} \cite{SLTUNET}, many of the modern and promising approaches, such as SignDiff \cite{SignDiff}, are not open-source. \\

    \item \textbf{Text-to-Speech (TTS)}. TTS is a well-explored research area. Hence, multiple open-source models are available to be used, such as Coqui TTS (XTTS-v2) \cite{XTTS}, Parler-TTS \cite{Parler-TTS_Git} \cite{Parler-TTS_Paper}, and MeloTTS \cite{MeloTTS}. \\

    \item \textbf{Speech-to-Text (STT, also known as ASR)}. This problem can be used both for offline mapping of speech-to-text and real-time captioning. Similar to TTS, the problem is well explored. Hence, multiple models are available to be used, such as Whisper \cite{Whisper} and wav2vec 2.0 \cite{Wav2Vec2.0}. \\

    \item \textbf{Image-to-Audio (ITA)}. This problem can be decomposed into two subtasks: Image-to-Text (ITT) and TTS. The classical models for image captioning are CLIP \cite{CLIP} and BLIP \cite{BLIP}. \\

    \item \textbf{Audio-to-Video (ATV)}. The ATV problem can be decomposed into two subtasks: SST and Text-to-Video (TTV). Converting text to video is currently an active research area, and we plan to try the following (but we are not limited to them) models: CogVideoX \cite{CogVideoX}, ModelScopeTTV \cite{ModelScopeT2V}. \\

    \item \textbf{Video-to-Audio (VTA)}. This problem can be decomposed into two subtasks: Video-to-Text (VTT) and TTS. VTT is the inverse of the TTV problem, yet it is not as popular. Hence, less research is done in this area. Nevertheless, several open-source models can be used, such as cogvlm2-llama3-caption \cite{CogVideoX}, and Video-LLaMA \cite{Video-LLaMA}. \\

    \item \textbf{Audio-to-Image (ATI)}. In this scenario, audio is converted to text, and then an image is created from the textual description. One of the effective ways to solve the Text-to-Image (TTI) problem is to apply stable diffusion \cite{StableDiffusion} models. \\
    
\end{itemize}

While the separate conversion problems (for example, TTS) do not require the involvement of agents, a more complex system can benefit from the use of Collective AI \cite{Collective_Intelligence}. Firstly, there is no unified model that can solve all the described modality conversion problems. Hence, agents with different roles and registered tools are useful for this purpose. Secondly, a user might not explicitly provide the expected output format. In this case, a multi-agent intelligent system will be capable of identifying the optimal output format based on collective reasoning. Meanwhile, a system not involving agents will face difficulties with this task. \\ 

\begin{table*}[t]
\caption{Agent Overview}
\begin{center}
\begin{tabular}{|p{2.2cm}|p{5.9cm}|p{2.9cm}||p{1.5cm}||p{2.8cm}|}
\hline
\textbf{Agent Name} & \textbf{Functionality} & \textbf{Input} & \textbf{Output} & \textbf{Default Model} \\
\hline
Interpreter Agent & Given a prompt, the agent determines the type of modality conversion desired by the user & StdIn & Category & ModCon-Task-Identifier\\
\hline
TTS Expert & Produces an audio transcription of the text & .txt, .pdf, .docx, StdIn & .wav & Tacotron 2 \cite{Tacotron2}\\
\hline
TTI Expert & Generates an image based on the text input & .txt, .pdf, .docx, StdIn & .png & Stable Diffusion V1-4 \cite{StableDiffusion}\\
\hline
STT Expert & Creates a text transcription of the audio file & .mp3, .mp4, .mpeg, .mpga, .m4a, .wav, .webm & .txt & Whisper \cite{Whisper}\\
\hline
ITT Expert & Generates text caption of the image & .png, .jpeg, .jpg & .txt & BLIP \cite{BLIP}\\
\hline
ATI Expert & Combines STT and TTI agents to generate an image based on the audio & .mp3, .mp4, .mpeg, .mpga, .m4a, .wav, .webm & .png & Whisper + BLIP\\
\hline
ITA Expert & Combines ITT and TTS agents to produce an audio description of the image & .png, .jpeg, .jpg & .wav & BLIP + Tacotron 2 \\
\hline
VTT Expert & Produces the text transcription of a video's audio component & .mp4, .webm & .txt & Whisper\\
\hline
\end{tabular}
\label{tab:agents}
\end{center}
\end{table*}

\begin{figure*}[htbp]
    \centering
    \includegraphics[width=0.8\linewidth]{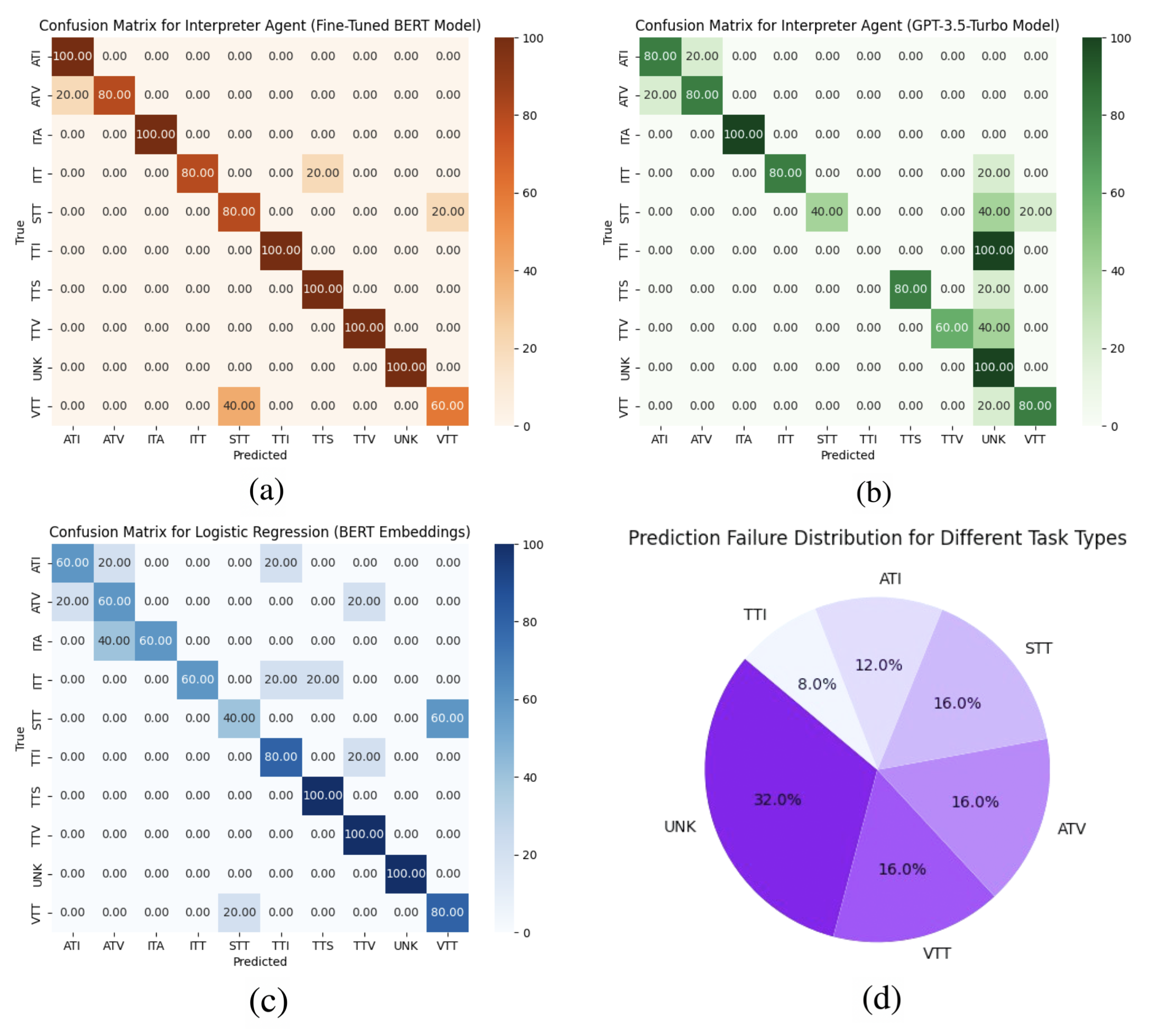}
    \caption{Comparison of Model Performance and Failure Distribution Across Task Types
        }
    \label{fig:confusion_matrices}
\end{figure*}

\section{Methodology}

\begin{table*}[t]
\caption{Comparison of LLM-Based Interpreter Agents}
\begin{center}
\begin{tabular}{c|cccccc}
\hline
\textbf{LLM Name} & \textbf{Accuracy} & \textbf{Precision} & \textbf{Recall} & \textbf{F1-Score} & \textbf{Failure Rate} \\
\hline
GLM-4-Flash \cite{ChatGLM} & 0.774 & 0.797 & 0.774 & 0.774 & 4/230 \\
Llama-3.1-70B-Instruct \cite{meta_llama31_2024} & 0.835 & 0.846 & 0.835 & 0.833 & 9/230 \\
GPT-3.5-Turbo \cite{GPT-3.5} & \textbf{0.865} & \textbf{0.880} & \textbf{0.865} & \textbf{0.865} & \textbf{1/230} \\
\hline
\end{tabular}
\label{tab:best_llm}
\end{center}
\end{table*}

The MATE system is defined as a set of agents $\mathcal{A}$:
\begin{equation}
\mathcal{A} = \{A_{IA}, A_{E_1}, ..., A_{E_7}\},
\end{equation}
where $A_{IA}$ is the interpreter agent and $\{A_{E_i}\}$ is the set of expert agents, capable of performing various modality conversion tasks. The system processes a user request $P_{user}$ by mapping it to an appropriate expert along with the input data $D_{In}$. The interpreter agent function $f_{IA}$ determines the modality conversion task type and is defined as:
\begin{equation}
f_{IA}(P_{user}) \rightarrow (A_{E_i}, D_{In}).
\end{equation}
After the task type is recognized, an appropriate expert executes its specialized modality conversion function $f_{E_i}$ on the input data to produce the final output $D_{Out}$:
\begin{equation}
f_{E_i}(D_{In}) \rightarrow D_{Out}.
\end{equation}
Fig. \ref{fig:architecture} shows the proposed MATE architecture. A user sends a request to the interpreter agent, describing the problem that needs to be solved (e.g., `convert text to audio', or `generate an audio description of the image'). Based on the prompt, the agent identifies the type of task and redirects this task to the appropriate agents. These agents are capable of solving specific modality conversion tasks, such as VTT, SST, etc. The output is a file of a desired format, and the path to that file is returned to the user. 

\subsection{Agents and User Interface}

Table \ref{tab:agents} provides a summary of the agents used in the MATE MAS. In total, eight agents were created using Microsoft's Autogen \cite{AutoGen} framework. The main agent (interpreter agent) is an LLM agent responsible for communicating with the user. An LLM used for the agent initialization can be both stored locally and called through the API. Upon receiving a prompt, it identifies the exact task that needs to be executed. Following this, the interpreter agent assigns the task to one of the seven experts for further execution. These experts use pre-defined models and functions to perform the modality adaptation tasks. Overall, agents are capable of solving the following modality conversion problems: TTS, TTI, ATI, STT, ITT, ITA, and VTT. In addition to agents' functionality description, Table \ref{tab:agents} shows their input and output format and the default models the agents are built on. \\

During initialization, the framework identifies a suitable folder for saving output files. This is achieved by searching for directories with specific names, such as `agents output', `data', etc. In case of the absence of such folders, MATE creates a new folder and returns its path to the user. After initialization, the system is capable of processing user requests. If the prompt is identified as irrelevant (UNK category), the system asks the user to re-enter the query so that the concrete modality conversion task can be specified. After the requirement is recognized, the user is asked to provide a path to the file that needs to be processed. In the cases of unknown, unexpected, or unrecognized file types, the user is asked to modify the file path accordingly. Next, the relevant agent completes the task and saves a new file in the pre-defined output directory. Finally, the path to this file is returned to the user, and the system is ready for further assistance.

\subsection{ModConTT Dataset Creation}

\begin{table*}[t]
\caption{Comparison of Task Classification Models}
\label{tab:best_model}
\centering
\setlength{\tabcolsep}{5mm}
\scalebox{0.9}{
\begin{tabular}{cc|ccccc}
 & \textbf{Model Name} & \textbf{Accuracy} & \textbf{Precision} & \textbf{Recall} & \textbf{F1-Score} \\
\hline
\multirow{4}{*}{TF-IDF} 
& Logistic Regression & 0.617 & 0.640 & 0.617 & 0.611 \\ 
& Random Forest & 0.650 & 0.659 & 0.650 & 0.629\\ 
& SVM & 0.583 & 0.605 & 0.583 & 0.580\\ 
& KNN & 0.517 & 0.524 & 0.517 & 0.514\\
& CatBoost & 0.617 & 0.600 & 0.617 & 0.582\\
\hline
\multirow{4}{*}{BERT} 
& Logistic Regression & \textbf{0.783} & \textbf{0.809} & \textbf{0.783} & \textbf{0.779} \\ 
& Random Forest & 0.600 & 0.623 & 0.600 & 0.584\\ 
& SVM & 0.600 & 0.662 & 0.600 & 0.588\\ 
& KNN & 0.583 & 0.626 & 0.583 & 0.584\\
& CatBoost & 0.667 & 0.664 & 0.667 & 0.646\\
\hline
\multirow{4}{*}{} 
& Interpreter Agent (GPT-3.5-Turbo) & \textbf{0.750} & \textbf{0.756} & \textbf{0.750} & \textbf{0.720}\\ 
& \textbf{Interpreter Agent (ModCon-Task-Identifier)} & \textbf{0.917} & \textbf{0.924} & \textbf{0.917} & \textbf{0.916}\\
\end{tabular}
}
\end{table*} 

Given the absence of existing datasets appropriate for training and evaluating models in modality conversion task type recognition, we constructed a diverse dataset (ModConTT) by leveraging LLMs to generate prompts aligned with various task types. These tasks include TTS, STT, ITT, ITA, VTT, TTI, ATI, TTV, ATV, and unknown (UNK). Prompts labeled with UNK refer to confusing, unrelated, or irrelevant requests from which the specific prompt type cannot be extracted. The dataset was made in two versions. The first version was created to identify the most accurate of the selected LLMs. These LLMs are available through API calls and are used for interpreter agent initialization. This version consists of 230 samples (20 for each category, except for the UNK class, which has 50 statements). The second version was used to train model-based task type classifiers from scratch. In total, there are 150 observations for the UNK category and 50 unique examples for each of the remaining task types, resulting in 600 prompts overall. Prompts in these two versions do not overlap. In ModConTT, each example includes a user prompt and the correct task type associated with this prompt. Existing large-scale LLMs were used to generate the prompts consistently and diversely so that the data is comprehensive for both training and evaluation. The dataset was designed to be applicable to various modality conversion use cases. Hence, the set of labels is larger than the number of modality adaptation tasks that the MATE system is capable of solving. For instance, MATE does not solve TTV tasks yet; however, TTV examples are present in ModConTT.

\subsection{Model Training and Evaluation}

In addition to using existing LLMs through API calls, five different classifiers (logistic regression, random forest, SVM, KNN, and CatBoost) were trained on the second version of the ModConTT dataset to identify the most suitable task type recognition model. The objective of these classifiers is to determine the most probable task type $t$ from a set of possible task types $T$ given $P_{user}$. The classification task is defined as:
\begin{equation}
\hat{t} = \arg\max_{t \in T} P(\,t\,|\,P_{user}; \theta_\mathcal{M}),
\end{equation}
where $\theta_\mathcal{M}$ represents the learned parameters of a given model $\mathcal{M}$, and $\hat{t}$ is the model prediction. Prompt embeddings were created using the TF-IDF method and the BERT \cite{BERT} model. Each of the classifiers was trained on both types of embeddings, resulting in 10 models overall. The data set was divided into train and test (90\% and 10\%, respectively) for all classifiers except CatBoost, where an additional 10\% of the train set observations were used as validation. All classifiers were trained with default hyperparameters, ensuring a fair comparison. Besides training classical machine learning models, we fine-tuned a BERT model on our train set. The number of early stopping rounds was set to be 5, and the total number of fine-tuning epochs was 6. The fine-tuned model (ModCon-Task-Identifier) was publicly \href{https://huggingface.co/AleksandrAlgazinov/ModCon-Task-Identifier}{released} on Hugging Face for reproducibility and further research, and all the trained classifiers are available at the project repository. Models were evaluated using standard classification metrics (accuracy, precision, recall, and F1-score). Precision, recall, and F1-score were averaged (using a weighted average strategy) across different classes to get a unique score for each model. In addition, the failure rate (FR) was calculated for LLM-powered interpreter agents, since they are not guaranteed to output a valid label. The failure rate is defined as the number of prompts where an agent failed to output a valid label from the pre-defined set, divided by the total number of prompts processed. Formally, given $N$ total prompts and a set of valid task labels $T$, the failure rate is defined as:
\begin{equation}
\text{FR} = \frac{\sum_{i=1}^{N} \mathbb{I}(\mathcal{M}(P_i) \notin T)}{N},
\end{equation}
where $\mathcal{M}(P_i)$ is the model's prediction for prompt $P_i$, and $\mathbb{I}(\cdot)$ is the indicator function.

\section{Results and Analysis}

\subsection{Evaluation of Interpreter Agents}

The performance of several existing LLMs (GLM-4-Flash \cite{ChatGLM}, Llama-3.1-70B-Instruct \cite{meta_llama31_2024}, and GPT-3.5-Turbo \cite{GPT-3.5}) was tested. GLM‑4‑Flash provides free API access, making it widely accessible for zero or low‑budget experimentation. Llama3.1‑70B, as an open‑source model, is relatively cheap for API calls while delivering strong performance, providing an adequate balance between resource consumption and output quality. The GPT models are widely regarded as a high standard in LLM quality, with their robust instruction-following and reliability for general NLP tasks. Hence, GPT-3.5-Turbo was selected as the third model. Importantly, the most advanced closed-source models, such as GPT-4 or Llama3.1‑405B, were avoided because their API usage is expensive, and the marginal performance gains might not justify the extra costs. Thus, our focus was on models that offer strong overall performance while remaining financially sustainable. Three separate interpreter agents were initialized under the configurations of these models, and their performance was evaluated based on the first version of the ModConTT dataset. \\

Table \ref{tab:best_llm} shows the comparison of the performance of these three models on the first version of the ModConTT dataset. GPT-3.5-Turbo outperformed both GLM-4-Flash and Llama3.1‑70B on all the metrics, achieving an accuracy of 0.865, precision of 0.880, recall of 0.865, F1-score of 0.865, and a failure rate of 1/230 ($\approx0.4$\%). Hence, the GPT-3.5-Turbo model is recommended for initializing the external LLM-powered interpreter agents. Fig. \ref{fig:confusion_matrices} (d) displays the distribution of prediction failures by task type. The highest failure rate was observed for the UNK class (32\%), since this category is the most common in the dataset. The second highest failure rates were detected for STT, ATV, and VTT categories (16\% for each one). Tasks involved in converting text and audio files to images were the simplest to recognize, accounting for 20\% of the overall failures.

\subsection{Evaluation of Task Classification Models}

Table \ref{tab:best_model} summarizes the performance of a variety of task classification models on the second version of the ModConTT dataset. Within the TF-IDF embeddings, random forest outperformed other ML-based classifiers, achieving an accuracy of 0.650 and an F1-score of 0.629. When using BERT representations, logistic regression achieved the best quality among non-LLM methods with an accuracy of 0.783 and an F1-score of 0.779. However, two of the top-3 performing systems are MAS-based interpreter agents. The GPT-3.5-Turbo agent achieved an accuracy of 0.750 and an F1-score of 0.720, while the fine-tuned BERT model (ModCon-Task-Identifier) agent attained the best overall classification results (accuracy 0.917 and F1-score 0.916), significantly outperforming other tested methods on all the metrics. This performance gap highlights the capacity of LLM-based agents to effectively capture complex task semantics. Hence, the results demonstrate that the use of MAS is justified for the modality conversion problem, as MAS architectures proved to be effective against ML models. \\ 

Fig. \ref{fig:confusion_matrices} shows confusion matrices for the three best-performing models: the ModCon-Task-Identifier agent (subfigure (a)), the logistic regression trained on BERT embeddings (subfigure (c)), and the GPT-3.5-Turbo agent (subfigure (b)). The ModCon-Task-Identifier agent is the most robust, consistently achieving an accuracy of 80\% or higher across task types (with the only exception for VTT, where the accuracy is 60\%). The overall performance of the GPT-3.5-Turbo model, conversely, is damaged by substantial errors in several classes (such as TTI and STT), making it less reliable. The logistic regression model displays more diverse error patterns across the classes; however, its overall performance remains relatively strong, establishing a high-quality baseline that can be achieved with a simpler machine learning architecture.

\section{Limitations and Future Work}

\subsection{Limitations of the Study}

While MATE represents an advancement in the AI for the accessibility domain, our research faces several limitations. \\

\begin{itemize}

    \item \textbf{External General Purpose Models}. While the choices of underlying models are flexible, our system relies on external tools for modality conversions. Hence, the MATE's performance can be compromised due to the model errors. In addition, most of the modality conversion models are trained for general-purpose applications and are not specifically designed for the accessibility domain, which could potentially lead to additional issues.

    \item \textbf{Lack of Video Generation Capabilities}. The current version of the framework does not support video generation models. Computational complexity and resource requirements of these models make them unsuitable for our real-time, efficient modality translation paradigm. Hence, although MATE offers extensive assistance for users in need, currently it is not a comprehensive solution for every possible scenario in the modality adaptation domain. For instance, TTSL and TTV cases are not handled by MATE.
    
\end{itemize}

\subsection{Future Work Directions}

In addition to limitations, several research directions are suggested for further enhancing MATE’s capabilities. \\

\begin{itemize}

    \item \textbf{Industrial Applications}. MATE can be directly integrated with digital hospital assistants to help improve patient care. MAS can support patients who have problems understanding medical information. For instance, using real-time TTS, MATE can turn a medical document into an audio representation, supporting people with vision impairment. In addition, MATE can be applied to the academic industry, helping disabled students get a high-quality education. Other possible application areas include transportation, retail, and entertainment.

    \item \textbf{Optimized TTV Models Integration}. As more efficient and lightweight TTV models are explored in the research community, the adaptability and scalability of MATE can be significantly enhanced, making it a viable solution for widespread use in resource-limited environments and across a wide range of industries. Hence, the future research can focus on optimizing MATE for domains where representing text and audio files as a video is crucial. For instance, a person with hearing difficulties can use ATV to follow cooking instructions through slow, visual demonstrations. Another example could be an individual using ATV to understand public transport directions as animated travel guides.
    
\end{itemize}

\section{Conclusion}

AI for accessibility remains a promising research direction, as existing technologies are usually designed for general-purpose use and fail to fully address the specific needs and preferences of users with disabilities. Even though MAS have proven to be powerful and effective in solving complex problems, their usage in the accessibility domain is limited. In this paper, we introduced MATE, the first open-source multi-agent framework created to offer comprehensive assistance to individuals with disabilities by performing modality conversions based on their needs. The core function of the framework is to convert data into a format that is accessible and understandable to the user. Additionally, using our custom dataset, we developed ModCon-Task-Identifier, a fine-tuned BERT model for modality task type recognition. Experiments show that the model achieves state-of-the-art performance on the ModConTT dataset, outperforming both existing LLMs and machine learning models trained from scratch. MATE's simplicity, flexibility, and light weight enable local system usage, ensuring privacy and convenience. With many potential applications, including healthcare, education, and transportation, MATE is expected to be a useful digital assistant to a wide range of users.

\bibliographystyle{ACM-Reference-Format}
\bibliography{references}

\end{document}